\documentclass[twocolumn,runningheads,natbib]{svjour2}
\bibpunct{[}{]}{;}{n}{}{,} 
\smartqed  
\usepackage{graphicx}
 \usepackage{mathptmx}      
\usepackage{latexsym}
\def\dj{d\kern-0.4em\char"16\kern-0.1em}
\journalname{Astrophysical Journal Letters}
\begin{document}

\title{Oscillatory behavior in the quiet Sun\\ observed with the New Solar Telescope}

\subtitle{High frequency oscillations observed with Rapid Dual Imager}

\titlerunning{Oscillatory behavior in the quiet sun observed with NST}        

\author{A.~An\dj {\,}i\'c     \and P.R.Goode \and J. Chae \and W. Cao \and K. Ahn \and V. Yurchyshyn \and V.Abramenko}

\authorrunning{An\dj {\,}i\'c  et. al} 

\institute{A.~An\dj {\,}i\'c \at
Big Bear Solar Observatory, 40386 North Shore Lane, BigBear City, CA 92314, USA \\
              Tel.: +1-909-8665791 ex 21\\
              Fax: +1-909-8664240\\
              \email{aandic@bbso.njit.edu}  \\
           \and P.R.Goode, J. Chae, W. Cao, K. Ahn, V. Yurchyshyn, V.Abramenko \at
Big Bear Solar Observatory, 40386 North Shore Lane, BigBear City, CA 92314, USA \\
\and  J. Chae, K. Ahn \at
Astronomy Program, Department of Physics and Astronomy,Seoul National University, Seoul 151-741, Korea \\}

\date{Received: date}

\maketitle

\begin{abstract}
Surface photometry of the quiet Sun has achieved an angular resolution of $0''.1$ with the New Solar Telescope at Big Bear Solar Observatory revealing that a disproportionate fraction of the oscillatory events appear above observed bright point-like structures. During the tracking of these structures, we noted that the more powerful oscillatory events are cospatial with them, indicating that observed flux tubes may be the source of many observed oscillatory events.

\keywords{photosphere \and oscillation}
\end{abstract}

\section{Introduction}
\label{intro}
When the excitation of oscillations of quiet-Sun (QS) regions is considered, one may think of the extensive research on acoustic oscillations being driven by accelerated down flows in the already dark intergranular lanes (Rimmele {\it et al.} 1995; Espagnet {\it et al.} 1996; Nordlund \& Stein 2001; An\dj {\,}i\'c 2007); that is, dark lanes that get locally precipitously cooler and darker representing a sharp accelerating down flow  followed by acoustic noise. However, the oscillations in the dark lanes could be more subtle with other types of oscillations being present as well. In particular, here we present observational evidence, using the New Solar Telescope (NST), for oscillations closely associated with the bright points (BPs) on the dark lanes. \par 
Some of the simulations of magnetohydrodynamics (V\"{o}gler et al. 2005 and references therein) showed the existence of the small flux tubes in the QS lanes, those were not observed in significant number until {\it HINODE} observations (Ceteno {\it et al.} 2007; Isobe {\it et al.} 2008; Lites {\it et al.} 2008).  \par

Recently, Khomenko {\it et al.} (2008) found that oscillatory events can, in addition, be caused by deep horizontal motions of the flux tube that generate a slow magnetic mode and a surface mode. They expect these modes to be efficiently transformed into slow acoustic modes in the $v_A < c_S$ atmosphere, where $c_S$ is the sound speed and $v_A$ the Alfv\'{e}n speed. Such a transformed mode should propagate along the field lines and might effectively deposit the energy of the  oscillatory driver into the chromosphere. Of course, this kind of the oscillatory generation can be seen only where flux tubes can be resolved.\par

Research so far established that Bright Points (BPs) are connected with magnetic flux concentrations, with a size of about $~$100 km, lifetimes of order of minutes and a tendency to form clusters if they are close to each other (Falchi {\it et al.} 1999; Berger \& Title, 1996; Berger {\it et al.}, 2004; Viticchie {\it et al.}, 2009).\par

In this work, we present the results of an analysis of the oscillations seen in a QS area that was recently observed using the 1.6 meter New Solar Telescope (NST) (Goode {\it et al.}, 2010) at Big Bear Solar Observatory (BBSO). The significantly greater spatial resolution of the NST makes possible observations providing a clearer picture of the mechanisms involved in QS dynamics. With increased resolution, we observed a plethora of the BPs in the QS area and noted cospatial with them, behavior of the oscillations that indicate that the flux tubes are an additional source of the oscillatory events observed in the QS. \par

\section{Observations}
\label{observations}

The data set used here was obtained on 29 July 2009. We used an optical setup at the Nasmyth focus of the NST of which the main  
components are a TiO broadband filter and a PCO.2000 camera. The TiO filter has a central wavelength at $705.68$ nm with a $1$ nm band pass. This filter has also a $1$nm leakage around $613$ nm with $25$\% of the transmission into the main spectral line. However, this does not jeopardize the interpretation of our observations. Broadband filters integrate in wavelength and average over the line and continuum intensities making the signal only weakly dependent on the properties of individual  
spectral lines. That is desirable in our case since the TiO line is very weak in the QS. Hence, we can assume that the height of  
formation in the data set is very close to that of the continuum.  
Use of the TiO filter allowed us to observe the solar photosphere at the diffraction limit of $0''.1$. We used a camera with 2048$\times$2048 pixels, where the pixel size is 7.4$\times$7.4~$\mu$m$^2$. The peak quantum  efficiency (QE) of a camera is at $500$ nm in the monochrome mode, yielding a QE 25\% at the observed wavelength. The QS was observed at disk center. \par

The data sequence consists of 120 bursts with 100 images in each burst. The exposure time for an individual frame was $10$ ms. The  
cadence of the data, between bursts, is $15$~s. The data were acquired in the morning with seeing levels that remained fairly constant throughout the observations. The images have a sampling of $0.037"$ per pixel. The CCD sampling was set to match the telescope's diffraction limited resolution in $G-band$, which caused a slight oversampling of the data.

\section{Data reduction and analysis methods}
\label{data}

The data obtained were speckle reconstructed based on the speckle masking method of \cite{luhe93} using the procedure and code  
described in \cite{woeger}. The short exposure times used for these data freeze out atmospheric distortions and at the same time capture the signals at high spatial frequencies (S\"{u}tterlin {\it et al.}, 2001); higher than studied here. \par

The cadence of our reconstructed data provided us with a Nyquist frequency of $67$~mHz. After speckle reduction, images were co-  
aligned using a Fourier co-aligning routine, which uses cross-correlation techniques and mean squared absolute deviation to provide sub-pixel co-alignment accuracy. However, we did not implement sub-pixel image shifting to avoid interpolation errors. Instead, the procedure was iterated $6$ times to achieve the best possible co-alignment. \par

We analyzed oscillations in the most reliable period range achievable with our data sets, from $87$ s to $700$ s. The decision to analyze only this range, even when our Nyquist frequency would allow access to shorter periods, came from camera induced noise that severely effected the signal for shorter periods. In order to strengthen reliability of our results we ignored the oscillations with the periods shorter than $87$ s.

Wavelet analysis of the light curves was performed using the Morlet wavelet:

\begin{equation}
\psi_0(t)= \pi^{-\frac{1}{4}}e^{i \omega_0 t} e^{-\frac{t^2}{2}},
\label{vaveleti2}
\end{equation}

\noindent where $\omega_0$ is a non-dimensional frequency and $t$ is a non-dimensional time parameter. The wavelet analysis code used in this work is based on the work of Torrence and Compo (1998), and our approach to automated wavelet analysis has been carried out previously, and presented in detail in Bloomfield {\it et al.} 2006.\par

To ensure that detected oscillations correspond to the real periodic motion, we applied several constraining criteria:\\

\begin{itemize}
\item {The light curve was tested against spurious detections of power that may be caused by Poisson noise, assuming that it is  
normally distributed and following a $\chi^2$ distribution with two degrees of freedom. A confidence level of $99$\% was set by  
multiplying the power in the background spectrum by the values of $\chi^2$ corresponding to the 99th percentile of the distribution, (Torrence \&Compo 1998; Mathiooudakis {\it et al.} 2003). }
\item{The light curve was compared with a large number (1500) of randomized time strings with identical count distributions. Comparing the value of power found in the input light curve with the number of times that the power transform of the randomized strings produced a peak of similar power. The probability of detecting non-periodic power was calculated for the peak power at each time-step (Banerjee {\it et al.}, 2001).}
\item{All oscillations of duration less than $1.3$ cycles were excluded in a comparison of the width of the peak in the wavelet  
power spectrum with the decorrelation time. This is done to distinguish between a spike in the data and a harmonic periodic  
component at the equivalent Fourier frequency, thus defining the oscillation lifetime at the period of each power maximum as the  
interval of time from when the power supersedes $95$\% significance, until it subsequently dips below $95$\% significance (McAteer {\it et al.} 2004). To obtain the number of cycles the lifetime is divided by the period.}
\item{All oscillations above a period of $701$ s were excluded as well, since they might be caused by edge effects due to the limited time sequence of our data. This highest credible period was calculated using the relationship between period of the oscillations and the length of our data set. We decided to ignore all periods that do not cover at least 2 cycles of the wave during our time series, in order to eliminate possible edge effects.}
\end{itemize}

We analyzed 42 BPs and tracked their displacement within the field of view. In our field of view, there was a plethora of observed BP-like structures. Since we do not have magnetic information, we choose the BPs using their evolution and size as determining factors. As criteria we used behavior, dimension and evolution of the BPs, which were already established in previous research (Berger {\it et al.}, 2004; Viticchie {\it et al.}, 2009). As the result, majority of the BPs analyzed here have an average size of $0''.2$, a lifetime around $5$~min and if they are close enough to each other, they tend to form clusters. To broaden our sample we included few of the BP-like structures that do not match these criteria. Two of them were present in images throughout our whole data set and 3 have very short lifetimes, being visible only for a minute. Each of them  was tracked using the NAVE methodology (\cite{jongchul08}). The NAVE method uses nonlinear affine velocity estimator to detect sub-pixel  motions, super-pixel motions and nonuniform motions.

\section{Results}
\label{results}

Previous works (Rimmele {\it et al.}, 1995; Espagnet {\it et al.}, 1996; Andic, 2007) noted oscillations are located mostly in intergranular lanes, i.e. darker parts of the observed granulation pattern. In the work by Andic (2007), $70$\% of all registered oscillatory power at a height of $200$ km above $\tau_{500}=1$ level was cospatial with the darker parts. However, the results here contradict those findings.  
We found that there is a significant contribution to the oscillatory power at the BPs observed in the QS area.\par
\begin{figure}
\centering
  \includegraphics[width=0.4\textwidth]{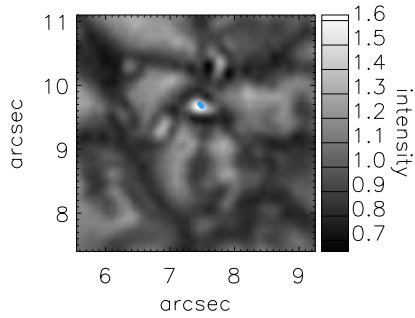}
\caption{The position of a detected oscillation in the field of view. The image shows detected power directly within a BP-like structure. Oscillations are represented with the blue contour. For this image, we took into consideration only oscillations with power above 80\% of maximum oscillatory power.}
\label{polo}       
\end{figure}

We analyzed separately the oscillations with the power above 80\% of the maximum oscillatory power. Our results showed that those  
oscillations tend to appear above the BP-like structures. Fig.\ref{polo} shows an example of one of those results. In the figure, this strong oscillation is presented with a blue contour, located at coordinates $(7''.5,9''.7)$, which encompasses only few pixels directly on the top of the BP structure. As long as this high oscillatory power was detectable in the data set, the structure did not showed any movement or visible evolution. These peaks of power lasted between $30$~s, and $120$~s. 

\begin{figure}
\centering
  \includegraphics[width=0.4\textwidth]{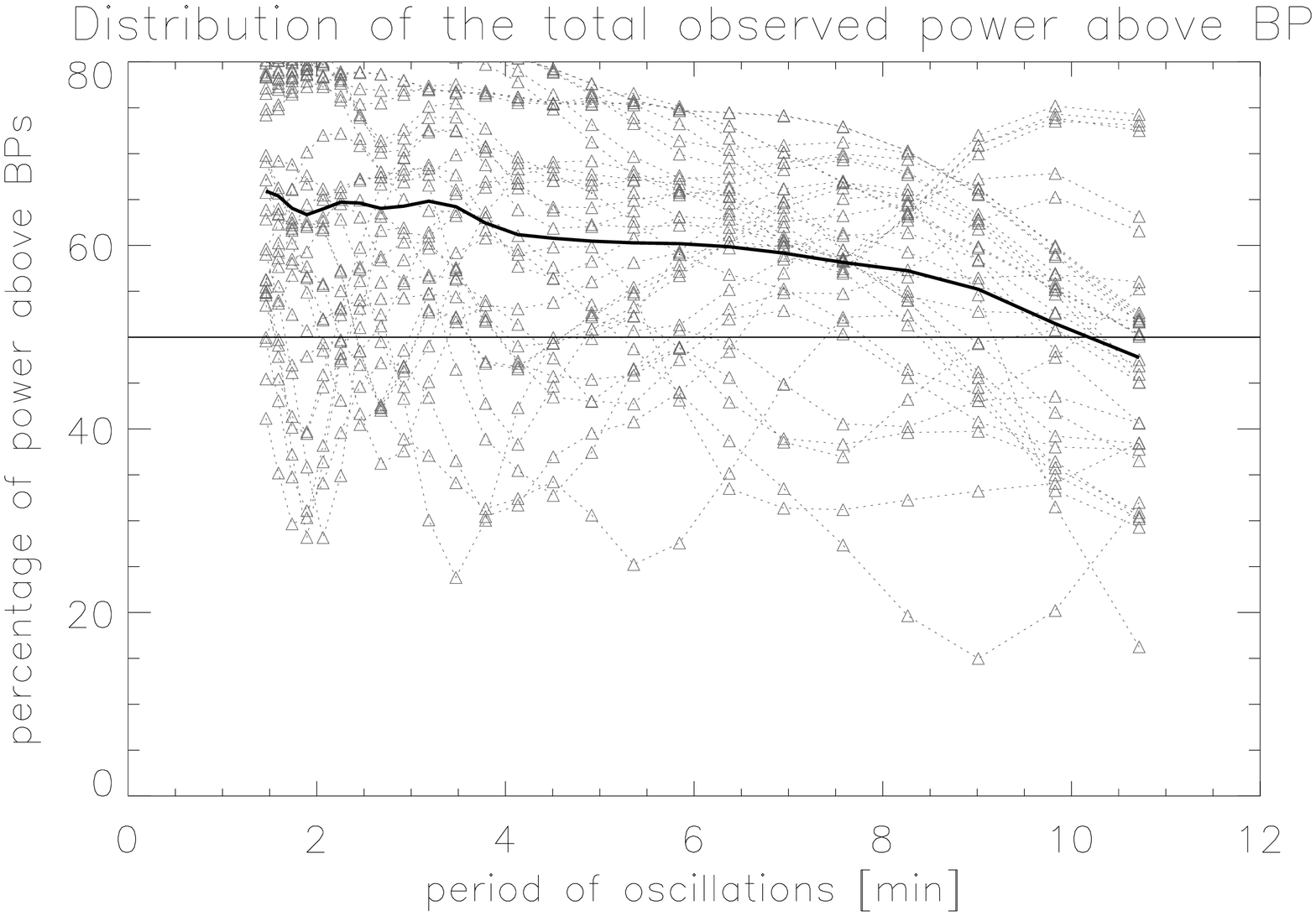}
\caption{The percentage of the oscillatory power in masked area that appears within the BP, covering the range of period analyzed, is presented for 42 BPs. The percentage is averaged over the time and space. The dotted lines with triangles present an average percentage of power that appeared within individual BPs. To illustrate the main trend, an average of all analyzed BPs is presented in thick solid line. And horizontal solid line marks the 50\% level.}
\label{distri}     
\end{figure}

Similar results were obtained when we included all registered power into the analysis. We took a sub-area of the image that surrounds the analyzed BP. The analyzed sub-area was $3''.5 \times 3''.5$, in average with the single BP was occupying around 3\% of the sub-area surface. We formed masks that isolated the analyzed BP in the sub-area, and an inverse mask that eliminated the BP in question. In Fig.\ref{distri}, for each analyzed BP, we showed the percentage of the power that appears within the BP across the period range. The percentage is averaged over the duration of our data and analyzed surface. The dotted lines represent the change in the percentage over the period range for each BP, the triangles present an average percentage for that particular period sub-range. To illustrate the main trend, an average is calculated from all analyzed BPs and it is presented as a thick solid line.  We note that for most period ranges the BPs tend to contribute significantly to the power in the sub-area. Most of the BPs had more than 50\% of the sub-areal power cospatial with them even though the BPs covered only about $3$\% of each sub-area.\par
It seems that there is a noticeable drop in contribution for longer periods. This might be a consequence of the BPs lifetime. An average lifetime of analyzed BPs was around $4.3$ min. Oscillations with a longer period than the lifetime cannot be observed at non-existent BPs.\par 
The wavelet analysis employs pixel by pixel curves, so if an oscillatory source moves from one pixel to another, the automated analysis will conclude that the oscillation ceased. Also due to the shift of wave train from one pixel to another, the wave train is interrupted and as such partially disregarded from our analysis because it does not fulfill the cycle conditions imposed on the data set (see section \ref{data}). The end result will be the detection of fewer oscillations. 
By tracking the detected source across the field of view, one can be assured that the source is always at the same pixel. This procedure ensures that the observed wave trains are not artificially interrupted but registered and analyzed in full. This procedure yields, as a result, a more accurate picture of oscillations in the QS.\par

We decided to follow a number (42) of BP-like structures and register the oscillations above them. We tracked the structures using the NAVE method. Then, using an information about the motion of the structure obtained with the NAVE method, we aligned the sub-images in such a way that a tracked structure was always in the same position.
As a comparison region for the amount of the emitted oscillations, we chose several sub-areas ($3''.5 \times 3''.5$) of true QS; meaning, we took care that these areas did not contained either a single BP or any other small structure during the whole duration of our time series. Those sub-areas contained only granules and the lanes that look empty at our resolution. This gave us a referent power of the emitted oscillations for comparison. 

\begin{figure}
\centering
  \includegraphics[width=0.35\textwidth]{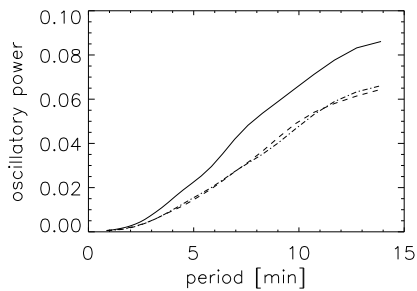}
\caption{The observed oscillatory power averaged over time and space in our data set. The solid line represent the oscillatory power observed within the 42 tracked BP structures. The dashed and doted-dashed lines represents the averaged oscillatory power above the area in which neither BP nor any small structure was detected during our time series, dashed line represent the power above granules, while doted-dashed line above the lanes.} 
\label{pow}
\end{figure}

By tracking BPs, we were able to observe the oscillatory power that is connected with the BPs and not their movement. In the Fig.\ref{pow}, we can see that oscillations detected within the BPs have more power than oscillations detected over the areas where neither a BP nor any small structure was observed during the time series. This apparent difference in the power may indicate different drivers for registered oscillations. The noticeable trend for longer period oscillations strengthen the conclusion that the drop in percentage from Fig.\ref{distri} for the same range is in part a consequence of the method of analysis.  

\begin{figure*}
\centering
  \includegraphics[width=0.48\textwidth]{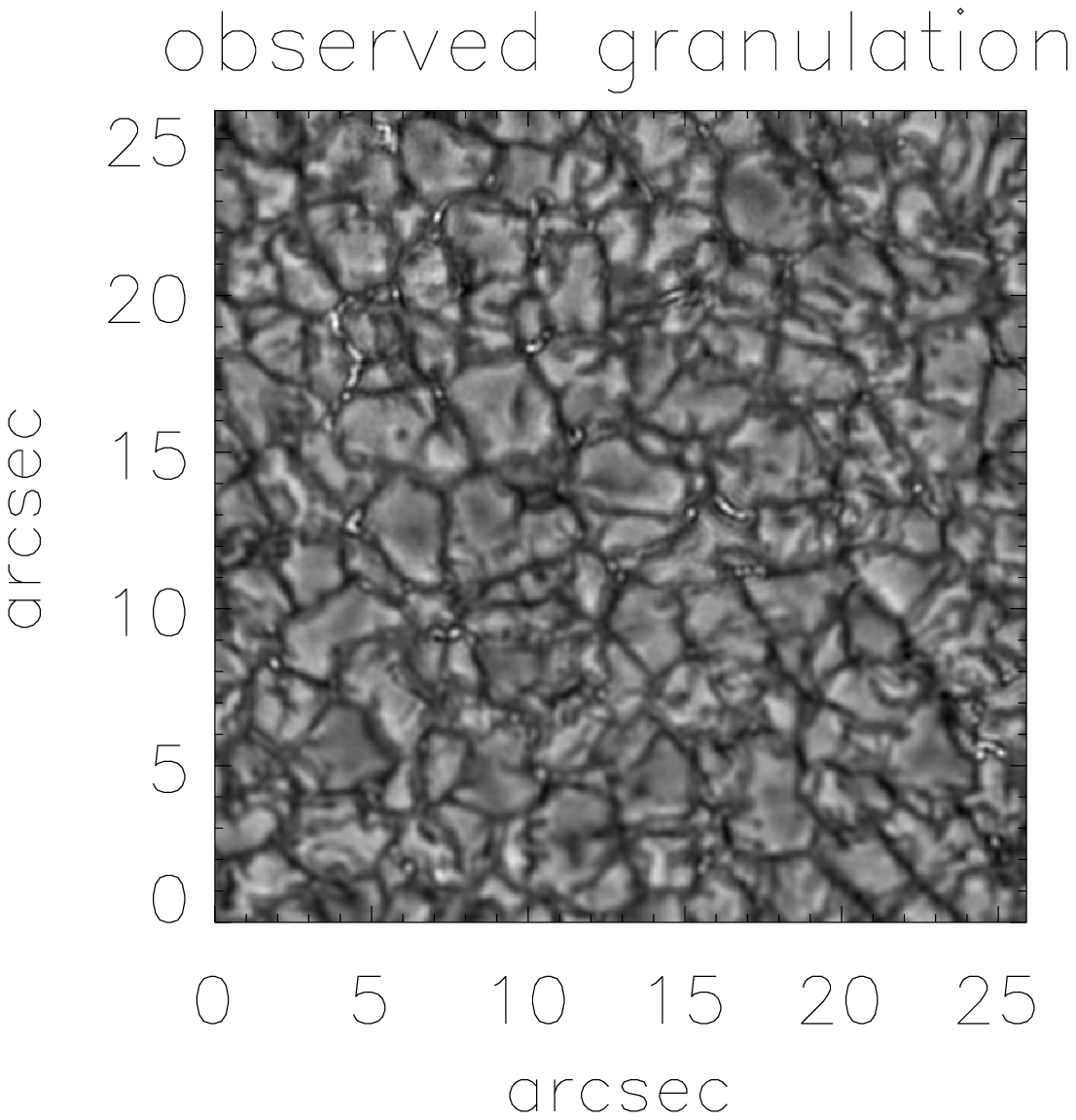}
   \includegraphics[width=0.48\textwidth]{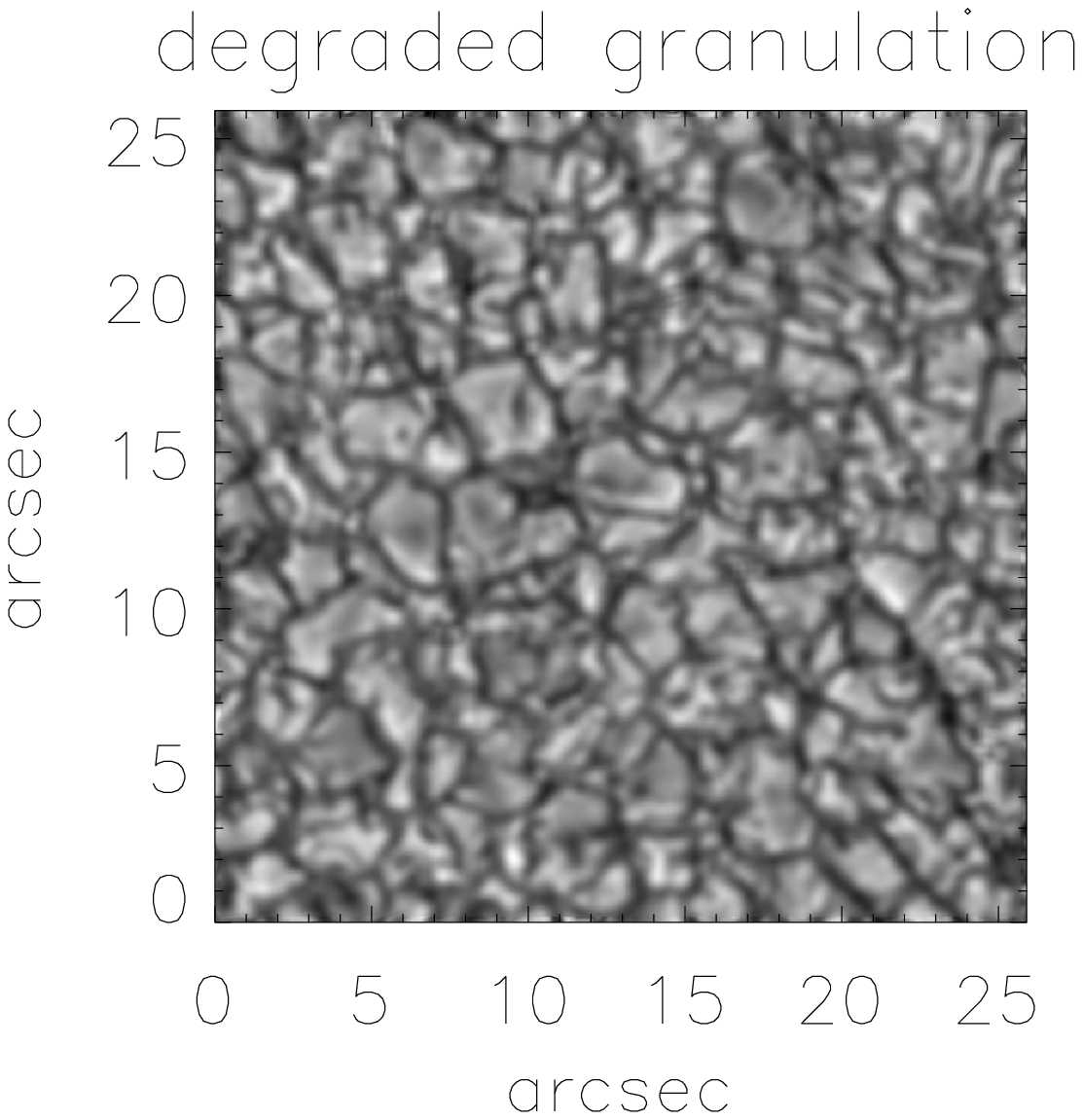}
\caption{The examples of the granular field. The left panel shows granulation as it is observed with the NST ($1.6$ m) at BBSO. The right panel shows how that same granulation would have looked if observed with the old, $0.6$ m telescope at BBSO. To make the right panel, the image on the left was degraded to match the resolution of the old telescope.}
 \label{granule}
\end{figure*}

With current resolution of the NST, we are capable of resolving much finer detail in the photosphere. However, if the same image is degraded to match the resolution of the old telescope in BBSO, we can see that only the most intense BPs structures remain visible, while the rest of them disappear in the lanes. This is illustrated in Fig.\ref{granule}.

The left panel of Fig.\ref{granule} presents a part of a random granulation image from our data set, while in the right panel is that same image degraded so that resolution of it corresponds to that of the old BBSO telescope. Hardly any BPs in the right panel can be detected in the left panel. \par

\section{Conclusions}
\label{conclusions}

When the proper motion of the BPs is removed, we can detect above them more powerful oscillations than above the areas in the field of view where neither a BP nor any small structure was detected. Work by Khomenko {\it et al.} (2008) argues that small flux tubes will create oscillations by their motions, and then conduct them upward. Our work indicates that this might be correct. \par

The higher resolution of the NST enabled us to detect a richer spectrum of BPs, and in turn yielded the conclusion that a fraction of the oscillatory power registered in QS lanes is essentially cospatial with BPs. Our plots showed a preference for the oscillations to appear above small structures in the granulation image.\par

We observe more oscillatory power within the BPs as compared to areas that do not contain any BPs. Since the observed extra power is co-spatial with the BPs, this might indicate that this extra power is generated by a driver connected with the BPs. One of the possible drivers is described in the work by Khomenko {\it et al.} (2008). Waves are generated by deep horizontal motions of the flux tubes; those motions generate a slow magnetic and a surface mode that are efficiently transformed into a slow acoustic mode inside of those flux tubes. However, we registered oscillations only at the height very close to $\tau_{500}=1$ level without any indication of the shock formation, so we can only speculate that this extra power is caused by the driver described in Khomenko {\it et al.} (2008). In favor of this speculation goes the simple deduction that if the all of the observed oscillations were generated without any contribution by BPs, then the averaged power over the compared areas should be similar in intensity. Since this is not a case, we have the implication that there is an additional oscillatory driver connected with BPs. \par

Most of the low intensity BPs were not detectable with earlier lower resolution observations and hence could not be connected to the oscillatory behavior, leading to the earlier conclusions that the all powerful oscillatory behavior in the QS is connected with the pure dark lanes (\cite{andic07}) and turbulent plasma flow. Our research implies that a part of oscillatory events we observed are connected with the previously unobserved plethora of the flux tubes in QS (Ceteno {\it et al.}, 2007; Isobe {\it et al.}, 2008; Lites {\it et al.}, 2008).  Due to the limitation of our data set we were unable to determine does these oscillations propagate upward and how much they contribute to the heating of the upper layers of the solar atmosphere.\par

Our main conclusion is that a part of the detected oscillations exists within the small magnetic flux tubes.  The observed additional oscillatory power is most likely generated by the additional driver closely connected with BPs. We expect to learn more about the nature of these oscillatory events with the installation of the newly developed instruments at NST that will allow us access to a high resolution Doppler maps and vector magnetograms at multiple heights in the solar atmosphere.\par

\begin{acknowledgements}
We would like to thank anonymous referee for the comments that helped to sharpen the some key point in this paper. We gratefully acknowledge support of NSF (ATM-0745744 and ATM-0847126), NASA (NNX08BA22G) and AFOSR (FA9550-09-1-0655). We are grateful to our knowledgable engineers for making possible for us to observe even during the final tuning of the telescope at Nasimyth focus.

\end{acknowledgements}



\end{document}